\documentclass [letterpaper,11pt]{article}

\usepackage{amsmath, amsthm}
\usepackage{amssymb}
\usepackage{graphicx}
\usepackage{verbatim}

\begin{document}

\title{\centering\Large{On Solutions for the Maximum Revenue Multi-item Auction \\
                               under Dominant-Strategy and Bayesian Implementations
                               }}
\date{}

\author{Andrew Chi-Chih Yao\thanks{Institute for Interdisciplinary Information Sciences, Tsinghua University, Beijing. This work was supported in part by the Danish National Research Foundation and the National Natural Science Foundation of China (under grant NSFC 61361136003).}}
\maketitle{}

\begin{abstract}
Very few exact solutions are known for the monopolist's $k$-item $n$-buyer maximum revenue problem with additive valuation in which $k, n >1$ and the buyers $i$ have independent private distributions $F^j_i$ on items $j$.  In this paper we derive exact formulas for the maximum revenue when $k=2$ and $F^j_i$ are any IID distributions on support of size 2, for both the dominant-strategy (DIC) and the Bayesian (BIC) implementations. The formulas lead to the simple characterization that, the two implementations have identical maximum revenue if and only if selling-separately is optimal for the distribution. Our results also give the first demonstration, in this setting, of revenue gaps between the two implementations.  For instance, if $k=n=2$ and $Pr\{X_F=1\}=Pr\{X_F=2\}=\frac{1}{2}$, then the maximum revenue in the Bayesian implementation exceeds that in the dominant-strategy by exactly $2\%$; the same gap exists for the continuous uniform distribution $X_F$ over $[a, a+1]\cup[2a, 2a+1]$ for all large $a$.
\end{abstract}

\newpage

\section{Introduction}

A monopolist wants to sell $k$ items to $n$ buyers with the aim to maximize revenue, where buyers $i$ have private (additive\footnote{Namely, for each buyer $i$, the valuation of a set $S$ of items is $\sum_{j\in S}t^j_i$, the sum of valuations for all the items in S.}) valuations $t^j_i$ of items $j$ described by independent distributions $F^j_i$ over the range $[0, \infty)$.  How should the optimal mechanisms be designed?

Myersons's classical paper [34] elegantly and completely solved the problem for the single item ($k=1$) setting.  For multiple items ($k>1$), the problem is much more complex with an extensive literature (see \emph{Related Work} below).  Much progress has been made, but many interesting questions remain open.

In this paper we focus on two such questions, both arising in connection with certain features of Myersons's solution.  Note that for $k=1$ Myerson's theory in many situations leads to explicit formulas for the optimal revenue, in addition to good intuitive understanding of how optimality arises.  For $k>1$, in the single buyer ($n=1$) case, there is a rich collection of sophisticated results (e.g., Manelli and Vincent [29], Hart and Nisan [23], Hart and Reny [25], Pavlov [35], Wang and Tang [40], Giannakopoulos and Koutsoupias [21][22],  Giannakopoulos [19][20]), where explicit expressions for optimal revenue are obtained for certain discrete and continuous distributions.  However, for $k>1$ and $n>1$, there do not seem to be any interesting results of this kind in the literature.

\noindent{\textbf{Question Q1.}}  For $k>1$ and $n>1$, can we obtain explicit expressions of the optimal revenue for interesting families of   distributions?

We turn to a second question. In auction theory, and more generally mechanism design theory, there are two standard versions of how the players' behavior is modeled, which translate into constraints on the classes of allowable mechanisms known respectively as \emph{dominant-strategy incentive-compatible} (DIC) and \emph{Bayesian incentive-compatible} (BIC) mechanisms. Formally, the BIC constraints look much weaker than the DIC constraints.  It is thus a remarkable feature of Myersons's theory for the single-item auctions that exactly the same maximum revenue is achieved by the BIC mechanisms and the DIC mechanisms.  Can this equivalence hold for $k>1$?

\noindent{\textbf{Question Q2.}}  For $k>1$, can Bayesian incentive-compatible (BIC) mechanisms ever produce strictly more revenue than the dominant-strategy incentive-compatible (DIC) mechanisms?

There is a substantial literature on the DIC versus BIC question (see\emph{ Related Work} below). When the independence condition on the distributions $F^j_i$ is dropped, then the answer to Question 2 is known.  Cremer and Mclean [15] showed BIC can generate unbounded more revenue than DIC, when $F^j_i$ are \emph{correlated across buyers} even for $k=1$.  Recently, Tang and Wang [39] showed in some instance with $k>1$, BIC can generate strictly more revenue than DIC, when $F^j_i$ are \emph{correlated across items}.  There are other examples (e.g. Gershkov et al. [18]) where DIC and BIC are shown to be inequivalent in revenue (and other attributes), but their models are farther away from our model under consideration here.

In this paper we address questions Q1 and Q2.  As a contribution in the direction of Q1, we derive exact formulas for the maximum revenue for both DIC and BIC implementations for $k=2$ and any $n>1$, where the 2n distributions $F^j_i$ are IID with a common $F$ of support size 2. As a by-product, these formulas give an answer to Q2, showing the BIC optimal revenue expression to be strictly greater than that of DIC for a broad range of parameters. In fact the formulas lead to the simple characterization that, the two implementations have identical maximum revenue if and only if selling-separately is optimal for the distribution. For instance, if $k=n=2$ and $Pr\{X_F=1\}=Pr\{X_F=2\}=\frac{1}{2}$, then the maximum revenue in the Bayesian implementation exceeds that in the dominant-strategy by exactly $2\%$.   A natural extension to the continuous case shows that the same $2\%$ gap holds for the uniform distribution over $[a, a+1]\cup [2a, 2a+1]$ as $a\rightarrow \infty$. We also remark that our result complements nicely a result in Yao [41] where the BIC maximum revenue is shown to be always upper bounded by a constant factor of the DIC maximum revenue.

\vskip 8pt
\noindent {\bf Related Work}

Much progress has been made on the computational aspects of multi-item auctions in models like the one discussed here. The intrinsic complexity of computing the optimal revenue has been investigated (e.g. [14][16]; efficient algorithms have been found in a variety of circumstances (e.g. [7][8][17]); furthermore, simple approximation mechanisms have been extensively studied in various environments (e.g. [1-3][6][9-13][23-28][37-39][41]).

The DIC versus BIC question falls in the domain of {\it Implementation Theory}, which is a central subject in mechanism design with a large literature (see e.g.[31-33][36]). Most related to our work, beyond references mentioned in previous paragraphs, are Manelli and Vincent [30], Gershkov et al. [18], which present equivalence results of DIC and BIC beyond revenue equivalence for models with one-dimensional types, and discuss the limitations of such equivalence results.

\section{Preliminaries}

\subsection{Basic Concepts}

Let $\mathcal{F}$ be a multi-dimensional distribution on $[0, \infty)^{nk}$.  Consider the $k$-item $n$-buyer auction problem where the valuation $n\times k$ matrix $t=(t_i^j)$ is drawn from $\mathcal{F}$.  Each buyer $i$ has $t_i\equiv(t_i^1, t_i^2, \cdots, t_i^k)$ as his valuations of the $k$ items. We also refer to $t_i$ as buyer $i$'s {\it type}, and $t$ as the {\it type profile} of the buyers (or \emph{profile} for short). For convenience, let $t_{-i}$ denote the valuations of all buyers except buyer $i$; that is, $t_{-i}=(t_{i'}\,|\,i'\neq i)$. Note that $t^j$, the $j$-th column of the matrix $t$, contains the valuations of all the buyers on item $j$.

A \emph{mechanism} $M$ specifies an \emph{allocation} $q(t)=(q_i^j(t)) \in [0, \infty)^{nk}$, where $q_i^j(t)$ denotes the probability that item $j$ is allocated to buyer $i$ when $t=(t_i^j)$ is reported as the type profile to $M$ by the buyers. We require that $\sum_{i=1}^n q_i^j(t)\leq 1$ for all $j$, so that the total probability of allocating item $j$ is at most $1$. $M$ also specifies a \emph{payment} $s_i(t)\in (-\infty, \infty)$ for buyer $i$.

The {\it utility} $u_i(t)$ for buyer $i$ is defined to be $t_i \cdot q_i(t) - s_i(t)$, where $t_i \cdot q_i(t)$ stands for the inner product $\sum_{j=1}^k t_i^j q_i^j (t)$. Let $u_i(t_i \leftarrow t_i', t_{-i})= t_i \cdot q_i (t'_i, t_{-i}) - s_i(t'_i, t_{-i})$, i.e. the utility buyer $i$ would obtain if he has type $t_i$ but reports to the seller as $t_i'$.  The {\it expected utility} ${\bar u}_i(t_i)$ for buyer $i$ is defined to be $E_{t_{-i}} (u(t_i , t_{-i}))$. Also let ${\bar u}_i(t_i \leftarrow t_i')=
 E_{t_{-i}} ({\bar u}_i(t_i \leftarrow t_i', t_{-i}))$.

The following formulas are well known:

\noindent{\bf Transfer Equations:} $u_i(t_i \leftarrow t_i', t_{-i})= u_i(t_i', t_{-i}) + (t_i -t_i')\cdot q_i(t_i', t_{-i})$ for all $t_i, t'_i, t_{-i}$.

\noindent{\bf Transfer Equations (Averaged form):} ${\bar u}_i(t_i \leftarrow t_i')= {\bar u}_i(t_i') + (t_i -t_i')\cdot {\bar q}_i(t_i')$ for all $t_i, t'_i$.

Two kinds of mechanisms have been widely studied, referred to as \emph{Dominant-strategy} and \emph{Bayesian} implementations as specified below.

\noindent{\textbf{A. Dominant-strategy Implementation}}

{\bf IR conditions:} $u_i(t) \geq 0$ for all $i$ and $t$.

{\bf DIC conditions:} $u_i(t_i, t_{-i}) \geq u_i(t_i \leftarrow t_i', t_{-i})$ for all $t_i, t_i'$ and $t_{-i}$,
or equivalently,

{\bf DIC conditions (Alternate):} $u_i(t_i, t_{-i}) - u_i(t'_i, t_{-i}) \geq (t_i -t_i')\cdot q_i(t_i', t_{-i})$ for all $t_i, t_i'$ and $t_{-i}$.

A mechanism is called {\it individually rational} ({\it IR})/{\it dominant-strategy incentive compatible} ({\it DIC}), if it satisfies the IR conditions/the DIC conditions, respectively.

\vskip 8pt
\noindent{\textbf{B. Bayesian Implementation}}

{\bf BIR conditions:} ${\bar u}_i(t_i) \geq 0$ for all $i$ and $t_i$.

{\bf BIC conditions:} ${\bar u}_i(t_i) \geq {\bar u}_i(t_i \leftarrow t_i')$ for all $t_i, t_i'$, or equivalently,

{\bf BIC conditions (Alternate):} ${\bar u}_i(t_i) - {\bar u}_i(t_i') \geq (t_i -t_i')\cdot {\bar q}_i(t_i')$ for all $t_i, t_i'$.

A mechanism is called {\it Bayesian individually rational} ({\it BIR})/{\it Bayesian incentive compatible} ({\it BIC}), if it satisfies the BIR conditions/the BIC conditions, respectively.

\vskip 8pt
Let $s(x)=\sum_{i=1}^n s_i(x)$ be the total payments received by the seller. For any mechanism $M$ on $\mathcal{F}$, let $M(\mathcal{F})= E_{x\sim \mathcal{F}}(s(x))$ be the (expected) revenue received by the seller from all buyers. The \emph{optimal} revenue is defined as $REV_D(\mathcal{F})=\sup _M M(\mathcal{F})$ when $M$ ranges over all the IR-DIC mechanisms. Similarly, in the Bayesian model, the optimal revenue is defined as $REV_B(\mathcal{F})=\sup _M M(\mathcal{F})$ where $M$ ranges over all the BIR-BIC mechanisms. As a benchmark for comparison, let $SREV(\mathcal{F})$ stand for the revenue yielded when each item is sold separately by using Myerson's optimal mechanism [34].

\vskip 8pt
\noindent{\textbf{Hierarchy Mechanism}}

 Consider an $n$-buyer 1-item auction.  A \emph{hierarchy allocation scheme} ${H}$ is specified by a mapping $Rank: T\rightarrow \mathcal{R}\cup\{\infty\}$.  Given a type $t\in (t_1, t_2, \ldots, t_n$), scheme ${H}$ allocates the item uniformly among the set of buyers $i$ with the smallest ranking.  If $Rank(t_i)=\infty$ for all $i$, then no allocation will be made to any buyer. For convenience, we also use the notation ${H}=[\tau_{11}, \ldots, \tau_{1 a_1};\tau_{21}, \ldots, \tau_{2 a_2}; \ldots;\tau_{\ell 1}, \ldots, \tau_{\ell a_\ell} ]$ with the understanding $Rank(\tau_{d m})=d$ for all $1\leq d\leq \ell$, $1\leq m\leq a_d$, and  $Rank(t)=\infty$ for any type $t$ not listed among $\tau_{d m}$.

In an $n$-buyer $k$-item auction, a \emph{hierarchy mechanism} $M$ uses an \emph{allocation function} specified by a $k$-tuple $\mathcal{H}=({H}^1, {H}^2, \cdots, {H}^k)$, where each ${H}^j$ is a hierarchy allocation scheme to be used for item $j$; also a utility function $u_i(t)$ for each buyer $i$ needs to be specified for $M$.  Note that the payment for buyer $i$ is determined by $s_i(t)=\sum_j q_i^j(t) t_i^j - u_i(t)$.

The concept of hierarchy mechanism was raised in Border [4][5] for one item, and later for multi-items in Cai et al.[7] in connection with Border's Theorem and efficiently computing optimal auctions.  Here we only need the concept as a convenient way to describe some of our proposed mechanisms.  More in-depth discussions of hierarchy mechanisms can be found in [4][5][7].

\section{Main Results}

In this paper, we solve for $REV_B(\mathcal{F})$ and $REV_B(\mathcal{F})$ in the $n$-buyer, $2$-item case when $\mathcal{F}$ consists of 2n IID's of a common $F$ with support size 2. Any such $\mathcal{F}$ can be specified by a 4-tuple $\delta=(n, p, a, b)$ where $n\geq 2$ is an integer, $0<p<1$, and $0\leq a<b$. Let $\mathcal{F}_\delta$ denote the valuation distribution for the $n$-buyer $2$-item auction, where the distributions $F^j_i$ for buyer $i$ and item $j$ are independent and identical (IID) copies of random variables $X$ defined by $Pr\{X=a\}=p$ and $Pr\{X=b\}=1-p$. Assuming additive valuation on items for each buyer, we are interested in determining $REV_D(\mathcal{F}_\delta)$ and $REV_B(\mathcal{F}_\delta)$, the maximum revenue achievable under IR-DIC and BIR-BIC, respectively, for distribution $\mathcal{F}_\delta$. We find two benchmarks relevant, $SREV(\mathcal{F}_\delta)$ and $s_b=2(1-{p^n})b$: the former is the revenue obtained by selling separately each item using Myerson's optimal mechanism [34]; the latter is the revenue by selling separately each item at price $b$.

\vskip 8pt
\noindent\textbf{Fact 1.} $SREV(\mathcal{F}_{\delta})=2\cdot\max\{(1-p^n)b, \ p^{n-1}a+(1-p^{n-1})b\}$.

\subsection{The Main Theorem}

For any real-valued function $G$, we use $G_+$ to denote the nonnegative function defined as  $G_+=\max\{G, 0\}$.

\noindent \textbf{Definition 1.}  Let $p_0=p^{2n}$, $p_1=2np^{2n-1}(1-p)$, and $p_2=2p^n\left(1-p^n-np^{n-1}(1-p)\right)$. Define
\begin{align*}
r_D(\delta)=2(1-{p^n})b\        &+\ p_0\Big[2a-\frac{1-p^2}{p^2}(b-a)\Big]_+\\
                                &+\ p_1\Big[a-\frac{1-p}{2p}(b-a)\Big]_+\\
                                &+\ p_2\Big[a-\frac{1-p}{p}(b-a)\Big]_+\\
r_B(\delta)=2(1- {p^n})b        &+\ p_0\Big[2a-\frac{1-p^2}{p^2}(b-a)\Big]_+\\
                                &+\ (p_1+p_2)\Big[a-\frac{1-p}{2p}(b-a)\Big]_+
\end{align*}

\noindent{\textbf{Main Theorem.}}\  $REV_B(\mathcal{F}_{\delta})=r_B(\delta)$ and  $REV_D(\mathcal{F}_{\delta})=r_D(\delta)$.
\vskip 8pt
\noindent{\textbf{Corollary 1.}\\
\noindent$(a)\   REV_B(\mathcal{F}_{\delta})=REV_D(\mathcal{F}_{\delta})= SREV(\mathcal{F}_{\delta}) \ \  \text{if}\ \  b\geq \frac{1+p}{1-p}a$,\\
\hspace*{18pt}$REV_B(\mathcal{F}_{\delta})>REV_D (\mathcal{F}_{\delta})> SREV (\mathcal{F}_{\delta}) \ \ \text{otherwise}$.\\
$(b)\   REV_D(\mathcal{F}_{\delta})=REV_B(\mathcal{F}_{\delta})\ \  \text{if and only if Selling-Separately is optimal}$.

\vskip 8pt

\noindent{\emph{Remark 1.}} For fixed $n, p, a$, the functions $r_B(\delta)$, $r_D(\delta)$ and $SREV(\mathcal{F}_{\delta})$ are each continuous, piecewise-linear functions in $b$ as shown in Figure 1. Breakpoints of linearity occur at $v_1=\frac{1+p^2}{1-p^2}a$, \ $v_2=\frac{1}{1-p}a$, and $v_3=\frac{1+p}{1-p}a$ \  along the $b$-axis. The two additive terms of $r_B(\delta)$ are 0 iff  $b\geq v_1$, and $b\geq v_3$, respectively. Similarly, the three additive terms of $r_D(\delta)$ are 0 iff  $b\geq v_1$, $b\geq v_3$, and $b\geq v_2$, respectively.

\begin{figure}
   \centering
    \includegraphics[scale=0.6]{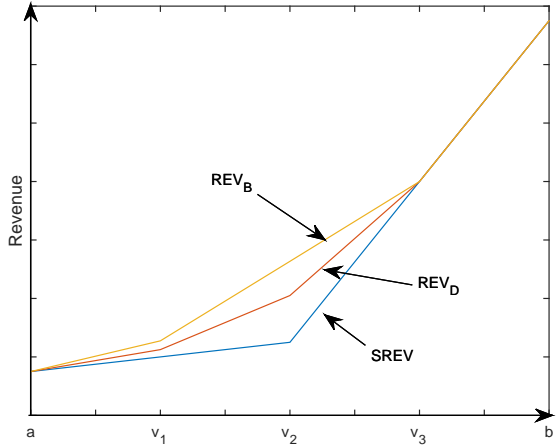}
    \caption{$REV_B, REV_D$, and $SREV$ as functions of $b$}
\end{figure}

\noindent{\emph{Remark 2.}}   The formula for $r_D(\delta)$  can be interpreted as follows (and likewise for $r_B(\delta)$).  The first term $2(1-{p^n})b$ is equal to $s_b$.  The three additive terms represent the extra revenue, beyond selling-separately at $b$, that can be gleaned from three specific subsets of profiles. (These subsets are defined as $S_0$, $S_1$, $S_2$ in Definition 5 later, with non-zero probability of occurrence $p_0, p_1, p_2$, respectively.)

\newpage
Corollary 1 can be derived as follows. When $b\geq v_3=\frac{1+p}{1-p}a$, all terms in square brackets in Definition 1 are 0, hence  $r_D(\delta)=r_B(\delta)=2(1-{p^n})b=s_b$. As by definition $s_b \leq SREV(\mathcal{F}_{\delta}) \leq r_D(\delta)$, we conclude $r_D(\delta)=r_B(\delta)=SREV(\mathcal{F}_{\delta})$ in this case. When $b< v_3$, we have $a-\frac{1-p}{2p}(b-a)>0$, implying
\begin{align*}
r_B(\delta)-r_D(\delta)&=p_2\Big(\Big[a-\frac{1-p}{2p}(b-a)\Big]_+  -\Big[a-\frac{1-p}{p}(b-a)\Big]_+\Big)> 0.
\end{align*}
 To compare $r_D(\delta)$ with $SREV (\mathcal{F}_{\delta})$ when $b< v_3$, notice that the two continuous piecewise-linear functions are equal when $b=a$ and $b=v_3$. It is easy to check from their formulas that, $r_D(\delta)$ strictly dominates $SREV (\mathcal{F}_{\delta})$ at both breakpoints $v_1$ and $v_2$ interior to $[a, v_3]$.  We conclude that $r_D(\delta)>SREV (\mathcal{F}_{\delta})$ over the entire interval $(a, v_3)$. Corollary 1(a) follows.  Corollary 1(b) follows immediately from 1(a).   \qed

\vskip 8pt
\noindent{\textbf{Example 1.}}
As an illustration, consider the case $\delta=(n, p, a, b)$ with $n=2, \ p=\frac{1}{2}, \ a=1, \ b=2$. The formulas in the Main Theorem tell us $REV_D(\mathcal{F}_{\delta})=\frac{1}{8}(3+11b)$ for $b\in[2,3)$, and $REV_B(\mathcal{F}_{\delta})=\frac{1}{16}(9+21b)$ for $b\in[\frac{5}{3},3)$. Thus for $b=2$ we have $REV_D(\mathcal{F}_\delta)=25/8=3.125$ and $REV_B(\mathcal{F}_\delta)=51/16=3.1875$, with a gap of $2\%$. Note that grand bundling yields only revenue $45/16$, while selling separately yields revenue 3,  both strictly less than $3.125$.

\subsection{Application to Continuous Distributions}

 The results in the Main Theorem have implications on the maximum revenue for continuous distributions if the latter can be well approximated by $\mathcal{F}_{\delta}$. As an application, let $\lambda>1$, $a>\frac{1}{\lambda-1}$, and let $\mathcal{F}=(F^j_i|\  1\leq i\leq n, 1\leq j \leq 2)$ be a distribution where $F^j_i=F$ are IID distributions with $support(X_F)=[a, a+1]\cup[\lambda a, \lambda a+1]$; let $p=Pr\{X_F\leq a+1\}$.  We can regard $F_\delta$, where $\delta=(n, p, 1, \lambda)$, as a normalized discrete approximation of $\mathcal{F}$.

\vskip 8pt
\noindent{\textbf{Corollary 2.}} Let $\delta=(n, p, 1, \lambda)$.  There exists a constant $C_\delta$ such that
\begin{align*}{|REV_Z(\mathcal{F})-r_Z(\delta)\cdot a|<C_\delta \ \ \text{for\ } Z\in \{D, B\}}.
\end{align*}

Corollary 2 is proved by an extension of our proof of the Main Theorem to the continuous setting (details omitted here).

\noindent{\emph{Remark 3.}} There are general high-precision approximation theorems in the literature (e.g. see[9][17][27][38]) connecting continuous and discrete distributions for the BIC maximum revenue auction. Our derivation of Corollary 2 does not rely on such general theorems.

We consider an illustrative example of Corollary 2 where $n=2$.  Let $\mathcal{G}_a=(F^j_i|\  i,j \in \{1, 2\})$, where $F^j_i=F$ are
IID distributions with $X_F$ uniformly distributed over $[a, a+1]\cup[2a, 2a+1]$.  According to Corollary 2, we have
$\displaystyle{\frac{1}{a}\lim_{a\rightarrow\infty} REV_Z(\mathcal{G}_a)=r_Z(\delta)}$\ \ for $Z\in \{D, B\}$ where  $\delta=(2, \frac{1}{2}, 1, 2)$.  More precise bounds for this example are given below.  Note that from Definition 1, one has $r_D(\delta)=\frac{25}{8}$ and $r_B(\delta)=\frac{51}{16}$, with a $2\%$ difference.

\vskip 8pt
\noindent{\textbf{Corollary 3.}} For $a\geq 20$, the BIC maximum revenue for $\mathcal{G}_a$ strictly exceeds its DIC maximum revenue.  In fact, we have for $a\geq 6$,
\begin{align*} \frac{25}{8}a &\leq REV_D(\mathcal{G}_{a})< \frac{25}{8}a+ \frac{5}{4}\\
              \frac{51}{16}a &\leq REV_B(\mathcal{G}_{a})< \frac{51}{16}a+ \frac{3}{2}.
\end{align*}

\subsection{Optimal Mechanisms}

The optimal revenue $r_D(\delta)$ and $r_B(\delta)$ stated in the Main Theorem can be realized, respectively, by the IR-DIC mechanism $M_{D, \delta}$ and the IR-BIC mechanism $M_{B, \delta}$ defined below. First, we name the characteristic functions for the three intervals where the individual terms of $r_D(\delta)$,  $r_B(\delta)$ are non-zero.

\vskip 8pt
\noindent \textbf{Definition 2.} Define
\begin{align*}
\begin{cases}
\alpha_{p,a,b}=1\ \  &\text{if} \ \ b< v_1, \ \ \ \text{and\ } 0 \ \  \text{otherwise}.\\
\beta_{p,a,b}=1\ \  &\text{if} \ \ b< v_3,     \ \ \ \text{and\ } 0 \ \  \text{otherwise}.\\
\gamma_{p,a,b}=1\ \   &\text{if} \ \ b< v_2,       \ \ \ \text{and\ } 0 \ \  \text{otherwise}.
\end{cases}
\end{align*}

The subscripts in $\alpha_{p,a,b},\  \gamma_{p,a,b},\  \beta_{p,a,b}$ can be dropped when $p,a, b$ are clear from the context. Note that, if desired, the formulas for $r_D(\delta)$ and $r_B(\delta)$ can be written using $\alpha, \beta, \gamma$ as multipliers in place of the notation $G_+\equiv\max\{G, 0\}$.

\noindent{\textbf{Definition 3.}} In what follows, the term {\it profile} refers to a profile in the support of $\mathcal{F}_{\delta}$, a {\it type} refers to a type in $\{ a, b\} \times \{a, b\}$. For any profile $t$ and $j\in \{1,2\}$, we say $t^j$ is \emph{cheap} if $t_i^j=a$ for all $1\leq i\leq n$ (we also say item $j$ is cheap); otherwise $t^j$ is \emph{non-cheap}. Call a profile $t$ \emph{1-cheap} if $t$ has exactly 1 cheap item.  We use $I(t)$ to denote the subset of buyers $i$ with $t_i\not =(a,a)$, that is, only excluding those who value both items at $a$. Note that, if $t$ is 1-cheap, then $|I(t)|$ is equal to the number of $b$'s in $t$ (and all appearing in the same column).

\vskip8pt
We now define mechanism $M_{D, \delta}$ and $M_{B, \delta}$ below, in the form of hierarchy mechanisms. First divide the range $(a, \infty)$ of $b$ into 4 subintervals:  $I_1=(a, v_1)$, $I_2=[v_1, v_2)$, $I_3=[v_2, v_3)$ and $I_4=[v_3, \infty)$.

\begin{description}
\item{\textbf{Mechanism $M_{D , \delta}$}:}
\item{\textbf{Case 1.}}  {$b\in I_1$. Use the allocation function $(H^1, H^2)$ where $H^1=[(b,b); (b,a); (a,b); (a,a)]$ and $H^2=[(b,b); (a,b); (b,a);(a,a)]$.}
\item{\textbf{Case 2.}} $b\in I_2$. Use the allocation function $(H^1, H^2)$ where $H^1=[(b,b); (b,a); (a,b)]$ and
$H^2=[(b,b); (a,b); (b,a)]$.
\item{\textbf{Case 3.}} $b\in I_3$. If $t=(t_i, t_{-i})$ with $t_{-i}$ being the lowest profile $(a,a)^{n-1}$, then offer items 1 and 2
to buyer $i$ as a bundle at price $a+b$. Otherwise, use the allocation function  $(H^1, H^2)$
where $H^1=[(b,b); (b,a)]$,  $H^2=[(b,b);(a,b)]$.
\item{\textbf{Case 4.}} $b\in I_4$.  Use the allocation function  $(H^1, H^2)$ where $H^1=[(b,b); (b,a)]$,  $H^2=[(b,b);(a,b)]$.
\end{description}
The payment of $M(\mathcal{F}_{\delta})$ is determined by the following utility function: for $1\leq i\leq n$, $t=(t_i, t_{-i})$,
\begin{align}u_i(t_i, t_{-i})=
\begin{cases}
           (b-a)\frac{\alpha}{n} \ \ &\text{if}\ \  t_i=(b,a)\ \text{or} \ (a,b), \  \text{and}\ \  t_{-i}=(a,a)^{n-1} \\
           (b-a)(\frac{\alpha}{n}+\beta) \ \ &\text{if}\ \  t_i=(b,b),\  \text{and}\ \  t_{-i}=(a,a)^{n-1} \\
           (b-a)\frac{\gamma}{1+|I(t_{-i})|} \ \ &\text{if}\ \  t_i=(b,b),\  t_{-i} \text{\ is 1-cheap} \\
           0  \ \ &\text{otherwise}.
\end{cases}
\end{align}

\noindent{\emph{Remark 4.}} Strictly speaking, $M_{D, \delta}$ in Case 3 is not a hierarchy mechanism.  We abuse the term slightly for convenience. We observe that when $b\in I_4$,  $M_{D, \delta}$ can be described as selling each item separately at price $b$ with a particular tie-breaking rule as dictated by the Case 4 allocation function $(H^1, H^2)$. When $b\in I_3$,  $M_{D, \delta}$ can be described as follows: if $t=(t_i, t_{-i})$ with $t_{-i}=(a,a)^{n-1}$, then offer items 1 and 2 to buyer $i$ as a bundle at price $a+b$; otherwise sell each item separately at price $b$ with a particular tie-breaking rule as dictated by the Case 3 allocation function $(H^1, H^2)$.

\begin{description}
\item{\textbf{Mechanism $M_{B,\delta}$}:}
\item{\textbf{Case 1.}}  $b\in I_1$: Use the allocation function as defined in Case 1 of $M_{D,\delta}$.
\item{\textbf{Case 2.}}  $b\in I_2\cup I_3$: Use the allocation function as defined in Case 2 of $M_{D,\delta}$.
\item{\textbf{Case 3.}}  $b\in I_4$: Define $M_{B,\delta}=M_{D,\delta}$.
\end{description}
In both Case 1 and 2, the payment is defined by the same utility function $u_i(t_i, t_{-i})$ as in $M_{D,\delta}$ with only one \emph{exception}: if $t_i=(b,b)$ and $t_{-i}$ is 1-cheap, then let
\begin{align*}
 u_i(t_i, t_{-i})=\frac{1}{2}(b-a)\frac{\beta}{1+|I(t_{-i})|}.
\end{align*}

\noindent\emph{Remark 5.}  In fact, $M_{B, \delta}(\mathcal{F}_\delta)$ is an IR-BIC mechanisms (not just BIR-BIC), as will be shown later.

\noindent{\textbf{Question.}} What is the difference between the allocation of $M_{D, \delta}$ and $M_{B, \delta}$, and how does $M_{B, \delta}$ manage to outperform $M_{D, \delta}$? We can gain some insights by looking at an example.

 Consider Example 1 introduced previously, with $\delta=(n, p, a, b)$ where $n=2$, $p=1/2$, $a=1$ and $b=2$. This $\delta$ falls under Case 3 of mechanism $M_{D , \delta}$ (and Case 2 of $M_{B , \delta}$, respectively). The two mechanisms are different only in the way they handle the following two sets of profiles: \\
 (A) Assume $t_1=t_2= (1,b)$ or $t_1=t_2= (b,1)$. Here $M_{B, \delta}$ offers each buyer $50\%$ of both items as a bundle at price $\frac{1}{2}(1+b)$, while $M_{D, \delta}$ offers each buyer $50\%$ of item 2 at price $\frac{b}{2}$; \\
 (B) Assume $i$ has type $(b, b)$ and the other buyer has type $(1, b)$ or $(b,1)$. Here $M_{B, \delta}$ offers buyer $i$ both items as a bundle at price $2b-\frac{1}{4}(b-1)$, while $M_{D, \delta}$ offers buyer $i$ both items as a bundle at price $2b$.

 Mechanism $M_{B, \delta}$ gets more payment than $M_{D, \delta}$ in situation A and gets less in situation B, but gains an overall improvement of $\frac{1}{16}$ over $M_{B, \delta}$. It is key to observe that mechanism $M_{B, \delta}$ \emph{violates} the DIC constraint $u_1((b,b), (1,b)) \geq u_1((1,b), (1,b)) + (b-1)q_1^1((1,b), (1,b))$.

\section{DIC Maximum Revenue}

In this section we give a proof outline of the Main Theorem for the dominant strategy implementation:

\noindent{\textbf{Theorem 1.}} Any IR-DIC mechanisms $M$ must satisfy $M(\mathcal{F}_{\delta})\leq r_D(\delta)$.

\noindent{\textbf{Theorem 2.}} $M_{D, \delta}$ is IR-DIC, and $M_{D, \delta}(\mathcal{F}_{\delta}) =  r_D(\delta)$.

We begin with a general discussion applicable to any mechanism. Let $M$ be a mechanism with allocation $q_i^j$ and utility $u_i$. We separate out the allocation of cheap items from non-cheap items. Thus, define ${q'}^{j}_i(t)=q^{j}_i(t)\eta^j(t)$ where $\eta^j(t)=1$ if item $j$ is cheap, and $\eta^j(t)=0$ if $j$ is non-cheap. Note that the welfare of the buyers from the allocation of cheap items is $a\cdot\sum_{i,j,t}Pr\{t\}{q'}^{j}_i(t)$, while the welfare from the non-cheap items is $\sum_{i,j,t}Pr\{t\}(1-\eta^j(t)) q_i^j(t)t_i^j$ which is at most $2(1-p^n)b$ (the revenue obtained by selling-separately at price $b$).  The well-known connection between revenue, welfare and utility leads to the following formula.

\noindent{\textbf{Basic Formula.}} For any mechanism $M$, we have \[ M(\mathcal{F}_{\delta})\leq 2(1-p^n)b+Qa-U, \] where
$ Q= \sum_{i,j,t}Pr\{t\}{q'}^{j}_i(t)$ and $ U    = \sum_{i,t}Pr\{t\}u_i(t)$.

\noindent{\textbf{Definition 4.}}  For any set $S$ of profiles, let $Q(S)=\sum_{t\in S}Pr\{t\} \sum_{i,j}{q'}_i^{j}(t)$, and
 $U(S)=\sum_{t\in S}Pr\{t\} \sum_{i}u_i(t)$.

To make use of the Basic Formula, we partition the profiles that can possibly contribute to the $Q$ term into three subsets $S_0$, $S_1$, $S_2$, and then use the IR-DIC Conditions to show that the $U$ term (utility obtained by buyers) is greater than a certain linear combination of $Q(S_0)$, $Q(S_1)$, $Q(S_2)$.  The Basic Formula then yields $r_D(\delta)$ as an upper bound to  $M(\mathcal{F}_{\delta})$.

\noindent{\textbf{Definition 5.}} Let $S_0=\{(a,a)^n\}$ be the set containing a single element, namely, the lowest profile.  Let $S_1$ be the set of 1-cheap profiles $t$ satisfying $|I(t)|=1$.  Let $S_2$ be the set of 1-cheap profiles $t$ satisfying $|I(t)|\geq 2$.

\vskip 8pt
\noindent\textbf{Fact 2.} ${q'}_i^{j}(t)=0$ for all $t\not\in S_0\cup S_1\cup S_2$.

\vskip 8pt
Recall that $p_0 = p^{2n}$, $p_1=2np^{2n-1}(1-p)$, $p_2= 2p^{n}(1-p^n-np^{n-1}(1-p))$. They have the following interpretation as can be easily verified.

\noindent\textbf{Fact 3.} $Pr \{ t \in S_\ell \}=p_\ell$ for $\ell \in \{ 0,1,2\}$,  where $t$ is distributed according to $\mathcal{F}_{\delta}$.

\vskip 8pt

\noindent{\textbf{Lemma 1.}}
\begin{equation}
Q(S_0)\leq 2p_0, Q(S_1)\leq p_1, Q(S_2)\leq p_2.
\end{equation}

\noindent{\it Proof.}  Any profile in $S_1$ or $S_2$ has exactly one cheap item, and the (only) profile in $S_0$ has two cheap items. Lemma 1 then follows from Fact 3.\qed

\noindent{\textbf{Lemma 2.}}
\begin{align}
M(\mathcal{F}_{\delta})\leq 2b(1-p^n)+a\sum_{0\leq \ell\leq 2}Q(S_\ell)-U.
\end{align}

\noindent{\it Proof.} It follows from Fact 2 that $Q=Q(S_0)+Q(S_1)+Q(S_2)$. Lemma 2 then follows from the Basic Formula.\qed

Lemma 1, 2 set the stage. We are ready to invoke the incentive compatibility requirements to prove Theorem 1. This setting is also useful in the next section when we prove the BIC part of the Main Theorem.

\subsection{Upper Bound to DIC Revenue}

We prove Theorem 1 in this subsection. The key is to prove the following proposition.

\noindent{\textbf{Proposition 1.}}  Any IR-DIC mechanism $M$ must satisfy the following inequality:
\begin{align}
U\geq (b-a)\big(\frac{1-p^2}{2p^2}Q(S_0)+\frac{1-p}{2p}Q(S_1)+\frac{1-p}{p}Q(S_2)\big).
\end{align}

We first show that Theorem 1 follows from Proposition 1.  It follows from Lemma 2 and Proposition 1 that, for any IR-DIC mechanism $M$, we have
\begin{align*}
M(\mathcal{F}_{\delta})\leq 2(1-p^n)b&+Q(S_0)\big[a-\frac{1-p^2}{2p^2}(b-a)\big]\\
                                     &+Q(S_1)\big[a-\frac{1-p}{2p}(b-a)\big]\\
                                     &+Q(S_2)\big[a-\frac{1-p}{p}(b-a)\big]\\
                       \leq 2(1-p^n)b&+Q(S_0)\big[a-\frac{1-p^2}{2p^2}(b-a)\big]_+\\
                                     &+Q(S_1)\big[a-\frac{1-p}{2p}(b-a)\big]_+\\
                                     &+Q(S_2)\big[a-\frac{1-p}{p}(b-a)\big]_+.
\end{align*}
With no negative terms, the above expression together with Lemma 1 immediately yield Theorem 1.  Thus to establish Theorem 1, it suffices to prove Proposition 1.

\noindent{\textbf{Definition 6.}}  For any $1\leq i, i' \leq n$, \\
 let $\tau_{i,i'}$ be the profile $t$ such that $t^{1}_i=t^{2}_{i'}=b$ and all other $t^j_\ell=a$; \\
 let $\tau_{i,0}$  be the profile $t$ with $t^{1}_i=b$ and all other $t^j_\ell=a$;\\
 let $\tau_{0,i'}$ be the profile $t$ with $t^{2}_{i'}=b$ and all other $t^j_\ell=a$;\\
 let $\tau_{0,0}=(a,a)^n$.

 \vskip 8pt
 \noindent{\textbf{Fact 4.}} $S_0=\{\tau_{0,0}\}$, $S_1=\{\tau_{i,0},\ \tau_{0,i} |\ 1\leq i \leq n\}$.

\vskip 8pt
\noindent{\textbf{Definition 7.}}  For any $t\in S_2$ and $1\leq i \leq n$, define $\tau_{t,i}$ as follows:
let item $j$ ($j\in \{1,2\}$) be the cheap item for $t$; define $\tau_{t,i}=t'$ where ${t'}^{j}_{i}=b$ and ${t'}^{j'}_{i'}=t^{j'}_{i'}$ for all other $(i', j')\not=(i,j) $.

\vskip 8pt
\noindent{\textbf{Definition 8.}}  Let $S'_1=\{ \tau_{i,i'}\, |\, 1 \leq i, i' \leq n \}$.  Let $S'_2 = \{  \tau_{t,i}   \,|\, t \in S_2, 1 \leq i \leq n \}$.

\noindent{\textbf{Fact 5.}} $S'_1,S'_2$ are disjoint sets of profiles containing no cheap items.

From Fact 5 and the IR Conditions, we have
\begin{align} U\geq U(S_1)+U(S'_1) + U(S'_2).
\end{align}

We now utilize the DIC-conditions to establish the following lemma relating the $U$ and $Q$ values on different types.

\noindent{\textbf{Lemma 3.}}
\begin{align} U(S_1)&\geq \frac{1-p}{p}((b-a)Q(S_0)+ 2U(S_0)),\\
U(S'_1)&\geq \frac{1-p}{2p}((b-a)Q(S_1)+ U(S_1)),\\
U(S'_2)&\geq \frac{1-p}{p}((b-a)Q(S_2)+ U(S_2)).
\end{align}

\noindent {\it Proof of Lemma 3.} The DIC-conditions require that, for all $t_i, t'_i, t_{-i}$,
\begin{align*} u_i(t_i, t_{-i})\geq u_i(t'_i, t_{-i}) + \sum_j (t_i^j-{t'}_i^{j})q_i^j(t'_i, t_{-i}).
\end{align*}
We only need a subset of these conditions where $t_i > t'_i$.  In such cases, we can use ${q'}_i^{j}$ instead of $q_i^{j}$ and write

\vskip 8pt
\noindent\textbf{{DIC-Conditions:}} For all $t_i > t'_i$ and any $t_{-i}$,
\begin{align} u_i(t_i, t_{-i})\geq u_i(t'_i, t_{-i}) + \sum_j (t_i^j-{t'}_i^{j}){q'}_i^{j}(t'_i, t_{-i}).
\end{align}
To prove Eq. 6, consider $t_i\in \{(b,a), (a,b)\}$, $t'_i=(a,a)$.  We have
\begin{align} u_i(\tau_{i,0})&=u_i((b,a), (a,a)^{n-1})\geq u_i(\tau_{0,0}) + (b-a){q'}_i^{1}(\tau_{0,0}), \nonumber \\
             \text{and}\ \ \   u_i(\tau_{0,i})&\geq u_i(\tau_{0,0}) + (b-a){q'}_i^{2}(\tau_{0,0}).
\end{align}
By Fact 4 we have
\begin{align*} U(S_1)&=\sum_{t\in S_1}Pr\{t\}\sum_{i'}u_{i'}(t)\\
                                       &=p^{2n-1}(1-p)\sum_{i}\big(\sum_{i'}u_{i'}(\tau_{i,0})+\sum_{i'}u_{i'}(\tau_{0,i})\big).
\end{align*}
Using Eq. 10 and the IR Conditions $u_{i'}(t)\geq 0$, we obtain
\begin{align*} U(S_1)&\geq p^{2n-1}(1-p)\big(\sum_{i}u_{i}(\tau_{i,0})+\sum_{i}u_{i}(\tau_{0,i})\big)\\
        &\geq p^{2n-1}(1-p)\sum_{i}\big(2u_{i}(\tau_{0,0})+(b-a)q_i^{'1}(\tau_{0,0})+(b-a)q_i^{'2}(\tau_{0,0})\big)\\
     &= \frac{1-p}{p}Pr\{\tau_{0,0}\} \big(2\sum_i u_{i}(\tau_{0,0})+(b-a)\sum_j\sum_i{q'}_i^{j}(\tau_{0,0})\big)\\
                     &=\frac{1-p}{p}(2U(S_0)+(b-a)Q(S_0)).
\end{align*}
This proves Eq. 6, the first inequality in the Lemma.

We now prove Eq. 7.  Write $S_1=S_1^L\cup S_1^R$ where $S_1^L=\{\tau_{0,i}|\ 1\leq i\leq n\}$ and $S_1^R=\{\tau_{i,0}|\ 1\leq i\leq n\}$. It suffices to prove for $x\in \{L,R\}$,
\begin{align} U(S'_1)\geq\frac{1-p}{p}((b-a)Q(S_1^x)+U(S_1^x)).
\end{align}
We prove Eq. 11 for $x=L$;  the case for $x=R$ is similar.
\begin{align} (b-a)Q(S_1^L)+U(S_1^L)&=\sum_{t\in S_1^L}Pr\{t\}\Big((b-a)\sum_j\sum_i {q'}_i^{j}(t)+ \sum_i u_i(t)\Big)\nonumber\\
                                    &=\sum_{t\in S_1^L}p^{2n-1}(1-p)\sum_i\big((b-a) {q'}_i^{1}(t)+ u_i(t)\big)\nonumber\\
                                    &=p^{2n-1}(1-p)\sum_{i'}\sum_i\big((b-a) {q'}_i^{1}(\tau_{0,i'})+ u_i(\tau_{0,i'})\big).
\end{align}

Now consider the DIC-Conditions (Eq. 9) for $(t_i, t_{-i})=\tau_{i,i'}$ and $(t_i', t_{-i})=\tau_{0,i'}$, which gives
\begin{align} u_i(\tau_{i,i'})\geq u_i(\tau_{0,i'})+(b-a)q_i^{'1}(\tau_{0,i'}).
\end{align}
From Eqs. 12 and 13, we obtain
\begin{align*} (b-a)Q(S_1^L)+U(S_1^L)&\leq p^{2n-1}(1-p)\sum_i\sum_{i'} u_i(\tau_{i,i'})\\
                                     &\leq \frac{p}{1-p}\sum_i\sum_{i'} Pr\{\tau_{i,i'}\}\sum_{i''}u_{i''}(\tau_{i,i'}) \\
                                     &= \frac{p}{1-p}\sum_{t\in S'_1} Pr\{t\}\sum_{i}u_i(t) \\
                                     &= \frac{p}{1-p}U( S'_1).
\end{align*}
This proves Eq. 11, thus completing the proof of Eq. 7.

We now prove Eq. 8, the third inequality of Lemma 3. By definition
\begin{align} (b-a)Q(S_2)+U(S_2)&= \sum_{t\in S_2}Pr\{t\}((b-a)\sum_{i,j} {q'}_i^{j}(t)+ \sum_i u_i(t)).
\end{align}
Now observe that the DIC-Condition Eq. 9 for $\tau_{t,i} \in S'_2$ and $t\in S_2$ implies\footnote{If $j$ is the cheap item in $t$, then $u_i(\tau_{t,i})\geq u_i(t)+(b-a) {q'}_i^{j}(t)$. However, ${q'}_i^{j}(t)=\sum_{j'} {q'}_i^{j'}(t)$ in this case, since ${q'}_i^{j'}(t)=0$ for $j'\not=j$.}
\begin{align} u_i(\tau_{t,i})\geq u_i(t)+(b-a)\sum_j {q'}_i^{j}(t)
\end{align}
From Eq. 14 and 15, we obtain
\begin{align*} (b-a)Q(S_2)+U(S_2)&\leq \sum_{t\in S_2}Pr\{t\}\sum_i u_i(\tau_{t,i})\\
                                     &=\frac{p}{1-p}\sum_{t\in S_2}\sum_i Pr\{\tau_{t,i}\}u_i(\tau_{t,i})\\
                                     &\leq \frac{p}{1-p}\sum_{t\in S'_2} Pr\{t\}\sum_{i''}u_i(t) \\
                                     &= \frac{p}{1-p}U( S'_2).
\end{align*}
This proves Eq. 8. We have completed the proof of the Lemma 3.  \qed

Proposition 1 can be straightforwardly derived from Lemma 3, Eq. 5, and the IR conditions $U(S_0), U(S_2) \geq 0$. This completes the proof of Proposition 1 and hence Theorem 1.

\subsection{Realizing DIC Revenue}

We turn to the proof of Theorem 2. We need to prove two statements.

\noindent {\bf Statement 1.} $M_{D, \delta}$ is IR and DIC;

\noindent {\bf Statement 2.} $M_{D, \delta}(\mathcal{F}_{\delta}) =  r_D(\delta)$.

The proof of Statement 1 is given in the Appendix. For the rest of this subsection, we prove Statement 2. Here is the top level view of the proof. To show that the upper bound on revenue from Theorem 1 can be achieved, we demonstrate that several critical inequalities involved in the upper bound proof can be replaced by equalities. First, for mechanism $M_{D, \delta}$, it can be verified that Eqs. 3, 4 now are equalities, while Eq. 2 is replaced by $Q(S_0)= 2\alpha p_0, Q(S_1)=\beta  p_1, Q(S_2)= \gamma p_2$. Combining these equalities gives us $M_{D, \delta}(\mathcal{F}_{\delta}) =  r_D(\delta)$. We now give the details.

\noindent{\textbf{Fact 6.}}  $u_i(t)=0$ for all $t\not\in S_1\cup S'_1\cup S'_2$ and all $i$. Thus, $U=U(S_1)+ U(S'_1)+U(S'_2)$.

 \noindent{\emph{Proof.}} From Eq. 1, we know that $u_i(t)\not =0$ may occur only when $t=(t_i, t_{-i})$ and one of the following is valid: (a) $t_{-i}=(a,a)^{n-1}$ and $t_i\not =(a,a)$;  (b) $t_{-i}$ is 1-cheap and $t_i=(b,b)$.  In case (a) we have $t\in S_1\cup S'_1$, and in case (b) we have $t\in S'_2$.  \qed

\noindent{\textbf{Fact 7.}}
\begin{align}\sum_{i,j} {q'}_i^{j}(t)=
\begin{cases}
           2\alpha \ \ &\text{if}\ \  t \in S_0 \\
          \beta \ \ &\text{if}\ \  t \in S_1 \\
           \gamma \ \ &\text{if}\ \  t \in S_2 \\
\end{cases}
\end{align}

\noindent{\it Proof.} For the (only) profile $t$ in $S_0$, the allocation function of $M_{D, \delta}$ specifies $\sum_{i,j} {q'}_i^{j}(t)= 2$ if $b < v_1$, and $0$ otherwise. Similarly, for any profile $t \in S_1$, $\sum_{i,j} {q'}_i^{j}(t)= 1$ if $b < v_3$, and $0$ otherwise; and for any profile $t \in S_2$, $\sum_{i,j} {q'}_i^{j}(t)= 1$ if $b < v_2$, and $0$ otherwise. This is exactly the assertion of Fact 7.\qed

\noindent{\textbf{Lemma 1'.}}
\begin{align}
Q(S_0)=2p_0 \alpha, Q(S_1)=p_1 \beta, Q(S_2)= p_2\gamma .
\end{align}

\noindent{\it Proof.} Follows immediately from Fact 3 and 7. \qed

\noindent{\textbf{Lemma 2'.}}
\begin{align*}
M_{D, \delta} (\mathcal{F}_{\delta}) = 2(1-p^n)b+ a\sum_{0\leq \ell\leq 2}Q(S_\ell)-(U(S_1)+ U(S'_1)+U(S'_2)).
\end{align*}

\noindent{\it Proof.} As under $M_{D, \delta}$ all the non-cheap items are allocated in full, the Basic Formula achieves equality, i.e.
$M_{D, \delta} (\mathcal{F}_{\delta}) = 2(1-p^n)b+Qa -U$. Also from Fact 2 we have $Q=Q(S_0)+Q(S_1)+Q(S_2)$, and from Fact 6 we have $U=U(S_1)+ U(S'_1)+U(S'_2)$. Lemma 2' follows.\qed

\noindent{\textbf{Lemma 3'.}}
\begin{align} U(S_1)&=(b-a)\frac{1-p}{p}Q(S_0),\\
U(S'_1)&=(b-a)\frac{1}{2}((\frac{1-p}{p})^2 Q(S_0)+(\frac{1-p}{p})Q(S_1)),\\
U(S'_2)&=(b-a)\frac{1-p}{p}Q(S_2).
\end{align}

\noindent{\emph{Proof.}}  To prove Eq. 18, note that using $M_{D, \delta}$'s utility definition in Eq. 1 we have
\begin{align*}
                           U(S_1)&= \sum_{t\in S_1}Pr\{t\}\sum_iu_i(t) \\
                                 &= \sum_{t\in S_1}Pr\{t\}\frac{\alpha}{n}(b-a) \\
                                 &=|S_1|p^{2n-1}(1-p)\frac{\alpha}{n}(b-a) \\
                                 &=2n p^{2n-1}(1-p)\frac{\alpha}{n}(b-a) \\
                                 &=(b-a)\frac{1-p}{p}Q(S_0),
\end{align*}
where we used Lemma 1' and Fact 3 in the last step. This proves Eq. 18.

To prove Eq. 19, note that for any $\tau_{i,i'} \in S'_1$, Eq. 1 implies
$\sum_{i''}u_{i''}(\tau_{i,i'})=(\frac{ \alpha}{n} + \beta) (b-a)$
if $i=i' $, and $0$ otherwise. Thus we have
\begin{align*}           U(S'_1)&= \sum_{t\in S'_1}Pr\{t\}\sum_{i''}u_{i''}(t) \\
                                 &= \sum_{i,i'} Pr\{\tau_{i,i'}\} \sum_{i''} u_{i''}(\tau_{i,i'})\\
                                 &= \sum_{i} Pr\{\tau_{i,i}\}  (\frac{\alpha}{n}+\beta)(b-a) \\
                                 &=p^{2n-2}(1-p)^2(\alpha+n\beta)(b-a).
 \end{align*}
 Making use of Lemma 1' and Fact 3, we obtain Eq. 19.

To prove Eq. 20, note that for any $t \in S_2, 1 \leq i, i' \leq n$ we have from Eq. 1
 \begin{align}u_{i'}(\tau_{t,i})=
\begin{cases}
           \frac{\gamma}{|I(t)|}(b-a) \ \ &\text{if}\ \ i'=i \in I(t)\\
           0  \ \ &\text{otherwise}.
\end{cases}
\end{align}
It follows that
\begin{align}U(S'_2)&=\sum_{t'\in S'_2}Pr\{t'\}\sum_{i'} u_{i'}(t')\nonumber\\
                    &= \sum_{t\in S_2}\ \sum_{i}\frac{1-p}{p}Pr\{t\}\sum_{i'} u_{i'}(\tau_{t,i})\nonumber\\
                    &=\frac{1-p}{p}\sum_{t\in S_2}Pr\{t\}\sum_{i\in I(t)}\frac{\gamma}{|I(t)|}(b-a)\nonumber\\
                    &=\frac{1-p}{p}\gamma(b-a)\sum_{t\in S_2}Pr\{t\}\nonumber\\
                    &= (b-a)\frac{1-p}{p} Q(S_2),
\end{align}
where we used Lemma 1' and Fact 3 in the last step. This proves Eq. 20. We have finished the proof of Lemma 3'.  \qed

 From Lemma 2' and 3', we have
 \begin{align*}
   M_{D, \delta}(\mathcal{F}_{\delta})=2(1-p^n)b &+Q(S_0)\big(a-(\frac{1-p}{p}+\frac{(1-p)^2}{2p^2})(b-a)\big)\\
                                                 &+Q(S_1)\big(a-\frac{1-p}{2p}(b-a)\big)\\
                                                 &+Q(S_2)\big(a-\frac{1-p}{p}(b-a)\big).
 \end{align*}
 Use Lemma 1' and simplify the above equation, we obtain
 \begin{align*}
    M_{D, \delta}(\mathcal{F}_{\delta})=r_D(\delta).
 \end{align*}
This proves Statement 2, and completes the proof of Theorem 2.

\section{BIC Maximum Revenue}

In this section we give a proof of the Main Theorem for the Bayesian implementation:

\noindent{\textbf{Theorem 3.}} Any BIR-BIC mechanisms $M$ must satisfy $M(\mathcal{F}_{\delta})\leq r_B(\delta)$.

\noindent{\textbf{Theorem 4}} $M_{B, \delta}$ is IR-BIC, and $M_{B, \delta}(\mathcal{F}_{\delta}) =  r_B(\delta)$.

\vskip 8pt

The proofs of Theorem 3 and 4 follows the same top-level outline as the proof of Theorem 1 and 2. Lemma 1 and 2 proved in Section 4 are valid for any mechanism $M$, and will also be the starting point for the BIC proof.

\subsection{Upper Bound to BIC Revenue}

We prove Theorem 3 in this subsection. The key is to prove the following proposition.

\noindent{\textbf{Proposition 2.}}  Any BIR-BIC mechanism $M$ must satisfy the following inequality:
\begin{align*}
U\geq (b-a)\big(\frac{1-p^2}{2p^2}Q(S_0)+\frac{1-p}{2p}(Q(S_1)+Q(S_2))\big).
\end{align*}

Theorem 3 can be derived from Lemma 1, 2 and Proposition 2 in exactly the same way as Theorem 1's derviation from Lemma 1, 2 and Proposition 1, and will not be repeated here. It remains to prove Proposition 2.

We use a subset of the BIR-BIC Conditions in our proof; these conditions are listed below for easy reference.

\noindent (a) BIR Condition: For each $i$,
\begin{align} \bar{u}_i(t_i)\geq 0 \ \text {where}\ \  t_i=(a,a).
\end{align}

\noindent (b) BIC Condition: For each $i$,
\begin{align}  \bar{u}_i(b,a)&\geq \bar{u}_i(a,a)+(b-a){\bar{q'}}_i^{1}(a,a).\\
               \bar{u}_i(a,b)&\geq \bar{u}_i(a,a)+(b-a){\bar{q'}}_i^{2}(a,a).\\
               \bar{u}_i(b,b)&\geq \bar{u}_i(a,b)+(b-a){\bar{q'}}_i^{1}(a,b).\\
               \bar{u}_i(b,b)&\geq \bar{u}_i(b,a)+(b-a){\bar{q'}}_i^{2}(b,a).
\end{align}
The plan is to use Eqs. 23-27 to obtain a lower bound on $U$ in terms of $Q(S_0)$, $Q(S_1)$ and $Q(S_2)$.

\vskip 8pt
\noindent{\textbf{Lemma 4.}}  For each $i$,
\begin{align*}\bar{u}_i(b,a) + \bar{u}_i(a,b) \geq (b-a)\sum_j{\bar{q'}}_i^{j}(a,a).
\end{align*}

\noindent{\emph{Proof.}} Immediate from  Eqs. 23-25. \qed

\vskip 8pt
\noindent{\textbf{Lemma 5.}}  For each $i$,
\begin{align*}\bar{u}_i(b,b) &\geq \frac{1}{2}(b-a)\sum_j{\bar{q'}}_i^{j}(a,a)\\
                                                          &+ \frac{1}{2}(b-a)\sum_j\big({\bar{q'}}_i^{j}(a,b)+{\bar{q'}}_i^{j}(b,a)\big).
\end{align*}

\noindent{\emph{Proof.}} Adding up  Eqs. 26 and 27, we obtain
\begin{align}\bar{u}_i(b,b)
&\geq \frac{1}{2}\big(\bar{u}_i(b,a) + \bar{u}_i(b,a)\big)\nonumber\\
& + \frac{1}{2}(b-a)\sum_j\big({\bar{q'}}_i^{j}(a,b)+{\bar{q'}}_i^{j}(b,a)\big),
\end{align}
where we have used the fact that ${q'}_i^{2}((a,b),t_{-i})={q'}_i^{1}((b,a),t_{-i})=0$ for all $t_{-i}$.  Lemma 5 now follows by using Lemma 4 on Eq. 28.  \qed

We now express $U$ as a convex combination of the left-hand sides of Eq 23, Lemma 4 and 5, and obtain a lower bound in terms of $Q(S_\ell)$:
\begin{align}U&=
\sum_i\sum_t Pr\{t\}u_i(t)\nonumber\\
&=p^2\sum_i \bar{u}_i(a,a)\nonumber\\
&+p(1-p)\sum_i \big(\bar{u}_i(b,a)+\bar{u}_i(a,b)\big)\nonumber\\
&+(1-p)^2\sum_i \bar{u}_i(b,b)\nonumber\\
&\geq C_1+C_2,
\end{align}
where
\begin{align}C_1&=(b-a)\big(p(1-p)+\frac{1}{2}(1-p)^2 \big)  \big[\sum_i\sum_{t_{-i}}Pr\{t_{-i}\}\sum_j {q'}_i^{j}((a,a),t_{-i})\big]\nonumber\\
                &=(b-a)\frac{1-p^2}{2}\sum_i\sum_{t_{-i}}Pr\{t_{-i}\}\sum_j {q'}_i^{j}((a,a),t_{-i}),\\
\text{and}\ \ \ \ \    C_2&=\frac{1}{2}(b-a)(1-p)^2\sum_i\sum_{t_{-i}}Pr\{t_{-i}\}\sum_j \big({q'}_i^{j}((a,b),t_{-i})+{q'}_i^{j}((b,a),t_{-i})\big).
\end{align}
Separating out the $t_{-i}=(a,a)^{n-1}$ term in Eq. 30, we obtain
\begin{align*}C_1&=(b-a)\frac{1-p^2}{2}\Big[\sum_i p^{2n-2}\sum_j {q'}_i^{j}(\tau_{00})\\
                  & \qquad\ +\sum_i \sum_{t_{-i}\not = (a,a)^{n-1}} Pr\{t_{-i}\}\sum_j {q'}_i^{j}((a,a),t_{-i})\Big]\\
                  &\geq (b-a)\frac{1-p^2}{2p^2} Pr\{\tau_{00}\}  \sum_{i,j} {q'}_i^{j}(\tau_{00})\\
                  &\qquad\ + (b-a)\frac{1-p}{2p} \sum_{\substack{t, i, t_i=(a,a)\\ t_{-i}\not = (a,a)^{n-1}}} Pr\{t\} \sum_j {q'}_i^{j}(t),\\
              C_2&=(b-a)\frac{1-p}{2p} \sum_{\substack{t,\  i\\ t_i\in \{(a,b), (b,a)\}}} Pr\{t\} \sum_j {q'}_i^{j}(t).
\end{align*}
It follows that
\begin{align*} C_1+C_2 &\geq (b-a)\frac{1-p^2}{2p^2}Q(S_0)\nonumber\\
                      &\ \ +(b-a)\frac{1-p}{2p}\sum_{t\not =\tau_{00}}\ \sum_{i,  t_i\not= (b,b)}Pr\{t\}\sum_j {q'}_i^{j}(t).
 \end{align*}
 Now, noting that $\sum_j {q'}_i^{j}(t)=0$ if $t_i= (b,b)$, we have
 \begin{align}\sum_{t\not =\tau_{00}}\ \sum_{i, t_i\not= (b,b)}Pr\{t\}\sum_j {q'}_i^{j}(t)
              &=\sum_{t\not =\tau_{00}}Pr\{t\}\sum_i\sum_j {q'}_i^{j}(t)\nonumber\\
              &=\sum_{t\in S_1\cup S_2}Pr\{t\}\sum_i\sum_j {q'}_i^{j}(t)\nonumber\\
               &= Q(S_1)+Q(S_2)
 \end{align}
 where we have used Fact 2.

 It follows from Eqs. 30-32 that
  \begin{align*} U\geq (b-a)\frac{1-p^2}{2p^2}Q(S_0)+ (b-a)\frac{1-p}{2p}(Q(S_1)+Q(S_2)).
  \end{align*}
  This proves Proposition 2, and completes the proof of Theorem 3.

\vskip 10pt
\subsection{Realizing BIC Revenue}

To prove Theorem 4, it suffices to prove the following two statements.

\noindent{\textbf{Statement 3.}}  $M_{B,\delta}$ is IR and BIC.

\noindent{\textbf{Statement 4.}}  $M_{B,\delta}(\mathcal{F}_{\delta})=r_B(\delta)$.

The proof of Statement 3 will be given in the Appendix. The rest of this subsection is devoted to the proof of Statement 4.  The statement is clearly true if $b \geq v_3$ (i.e. {\it Case 3} in the definition of $M_{B,\delta}$),  since in this case by definition $r_B(\delta)= r_D(\delta)$, $M_{B}(\mathcal{F}_{\delta})=M_{D}(\mathcal{F}_{\delta})$, and Theorem 2 has established $M_{D}(\mathcal{F}_{\delta})=r_D(\delta)$.  Thus we can assume $b\in (a, v_3)$ (i.e. Case 1 or 2).  Note that in this situation $\beta=1$ and $\alpha\in \{0, 1\}$.

The proof follows essentially the same outline as the proof of Statement 2 in Section 4.2.  Fact 2, 3, 6 remain true; Fact 7, Lemma 1', 2' are modified  to the following.

\vskip 8pt
\noindent{\textbf{Fact 8.}}
\begin{align*}\sum_{i,j}{q'}_i^{j}(t)=
 \begin{cases}2\alpha \ \ &\text{if}\ \ t\in S_0\\
              \beta \ \ &\text{if}\ \ t\in S_1\cup S_2.
 \end{cases}
\end{align*}

\noindent{\textbf{Lemma 1''.}}  \begin{align*}
                            Q(S_0)&=2p_0\alpha, \\
                            Q(S_1)&=p_1\beta, \\
                            Q(S_2)&=p_2\beta.
                           \end{align*}

\noindent{\textbf{Lemma 2''.}} $M_{B,\delta}(\mathcal{F}_{\delta})=2b(1-p^n)+a\sum_{0\leq \ell\leq 2}Q(S_\ell)-(U(S_1)+U(S'_1)+U(S'_2))$.

All the above statements are straightforward to prove. Finally, Lemma 3' is modified to the following:

\noindent{\textbf{Lemma 3''.}}
\begin{align}
U(S_1)&=(b-a)\frac{1-p}{p}Q(S_0),\\
U(S'_1)&=(b-a)\frac{1}{2}((\frac{1-p}{p})^2 Q(S_0)+(\frac{1-p}{p})Q(S_1)),\\
U(S'_2)&=(b-a)\frac{1-p}{2p}Q(S_2).
\end{align}

The proof of Eqs. 33-34 is exactly the same as in the proof of Eqs. 18-19 in Lemma 3'. The proof of Eq. 35 is also
  similar to the proof of Eq. 20 in Lemma 3', except that
Eq. 21 should be replaced by
 \begin{align}u_{i'}(\tau_{t,i})=
   \begin{cases}\frac{\beta}{2|I(t)|}(b-a) &\qquad \text{if}\ \ i'=i \in I(t)\\
                 0                         &\qquad \text{otherwise.}
   \end{cases}
 \end{align}
Proceeding as before, we obtain instead of Eq. 22,
  \begin{align*}U(S'_2)&=\frac{1-p}{2p}\beta(b-a)\sum_{t\in S_2}Pr\{t\}\\
              &=(b-a)\frac{1-p}{2p}Q(S_2).
\end{align*}
This proves Eq. 35, and completes the proof of Lemma 3''.   \qed

 It follows from Lemmas 2'', 3'' that
  \begin{align*}
   M_{B, \delta}(\mathcal{F}_{\delta})=2(1-p^n)b &+Q(S_0)\big(a-(\frac{1-p}{p}+\frac{(1-p)^2}{2p^2})(b-a)\big)\\
                                                 &+Q(S_1)\big(a-\frac{1-p}{2p}(b-a)\big)\\
                                                 &+Q(S_2)\big(a-\frac{1-p}{2p}(b-a)\big).
 \end{align*}
 Use Lemma $1''$ and simplify, we obtain
 \begin{align*}
    M_{B, \delta}(\mathcal{F}_{\delta})=r_B(\delta).
 \end{align*}
This proves Statement 4, and completes the proof of Theorem 4.

\newpage
\noindent{\textbf{\Large{Appendix}}

\vskip 6pt

\noindent{\textbf{A. Proof of Statement 1}}

We prove Statement 1 from Section 4.2. Without loss of generality (and making it easier to read), we normalize the value so that $a=1$ and $b>1$ for Appendix A and B. (The case $a=0$ is trivial, and can easily be checked separately.)

\vskip 8pt
\noindent{\textbf{Statement 1.}}  $M_{D, \delta}$ is IR and DIC.

\vskip 8pt
\noindent{\emph{Proof}.}  Mechanism $M_{D, \delta}$ is obviously IR, as the utility as defined by Eq 1 (main text) is always non-negative.  $M_{D, \delta}$ is also clearly DIC if $b\in[v_2, \infty)$, as $M_{D, \delta}$ can be defined in the form of a menu for each buyer (see Remark 4 in the main text).  Therefore, in our proof, we can assume that $b\in (1,v_2)$ and, in this situation, the parameters satisfy $\alpha\in \{0,1\}$ and $\gamma=\beta=1$.

The DIC Conditions can be written as: for all $t_i$, $t'_i$, $t_{-i}$,
\begin{align*} (t_i - t'_i)\cdot q_i(t_i, t_{-i})\geq u_i(t_i, t_{-i})-u_i(t'_i, t_{-i}) \geq (t_i - t'_i)\cdot q_i(t'_i, t_{-i}). \tag{A1}
\end{align*}
Note that Eq. (A1) is unchanged if we swap $t_i$ and $t'_i$.

As $M_{D, \delta}$ is symmetric in items and buyers, to show that $M_{D, \delta}$ is DIC, it is sufficient to prove the following subset of inequalities among (A1) for all $t_{-1}$:
\begin{align*} (0, b-1)\cdot q_1((1,b), t_{-1})&\geq u_1((1,b), t_{-1})-u_1((1,1), t_{-1})\\
                                               &\geq (0, b-1)\cdot q_1((1,1), t_{-1})  \tag{A2} \\
               (b-1, 0)\cdot q_1((b,b), t_{-1})&\geq u_1((b,b), t_{-1})- u_1((1,b), t_{-1})\\
                                               &\geq (b-1, 0)\cdot q_1((1,b), t_{-1}) \tag{A3} \\
             (b-1, b-1)\cdot q_1((b,b), t_{-1})&\geq u_1((b,b), t_{-1})- u_1((1,1), t_{-1})\\
                                                 &\geq (b-1, b-1)\cdot q_1((1,1), t_{-1})\tag{A4} \\
               (b-1, -(b-1))\cdot q_1((b,1), t_{-1})&\geq u_1((b,1), t_{-1})- u_1((1,b), t_{-1})\\
                                                 &\geq (b-1, -(b-1))\cdot q_1((1,b), t_{-1}). \tag{A5}
\end{align*}

\noindent{\textbf{Case A.}} $t_{-1}=(1^{n-1}, 1^{n-1})$.

By the definition of $M_{D, \delta}$, it is easy to verify that
\begin{align*} q_1(t_1, t_{-1})=
\begin{cases}
(\frac{\alpha}{n}, \frac{\alpha}{n})\ \  &\text{if} \ \ t_1=(1,1) \\
(1,1)\ \  &\text{if} \ \ t_1\not=(1,1);
\end{cases}
\end{align*}
\begin{align*}u_1(t_1, t_{-1})=
\begin{cases}
0 \ \ &\text{if}\ \  t_1=(1,1)\\
(b-1)\frac{\alpha}{n}\ \  &\text{if} \ \ t_1=(1,b)\  \text{or} \  (b,1) \\
(b-1)(\frac{\alpha}{n}+1)\ \  &\text{if} \ \ t_1= (b,b).
\end{cases}
\end{align*}

Thus, to verify A2-A5 is the same as to verify the following inequalities, which are obviously true:
\begin{align*}b-1&\geq (b-1)\frac{\alpha}{n}\geq (b-1)\frac{\alpha}{n},\\
                                b-1&\geq b-1\geq b-1,\\
2(b-1)&\geq (b-1)(\frac{\alpha}{n}+1)\geq (b-1)\cdot 2\frac{\alpha}{n},\\
0&\geq 0 \geq 0.
\end{align*}

\noindent{\textbf{Case B.}} $t_{-1}=(1^{n-1}, b^m 1^{n-m-1})$, $m\geq 1$.

By the definition of $M_{D, \delta}$, it is easy to verify that
\begin{align*} q_1(t_1, t_{-1})=
\begin{cases}
(0,0) \ \ \ \ &\text{if} \ \ t_1=(1,1) \\
(\frac{1}{1+m}, \frac{1}{1+m})\ \  &\text{if} \ \ t_1=(1,b) \\
(1,0)\ \  &\text{if} \ \ t_1=  (b,1)\\
(1,1)\ \  &\text{if} \ \ t_1=  (b,b);
\end{cases}
\end{align*}
\begin{align*}
u_1(t_1, t_{-1})=
\begin{cases}
0 \ \ &\text{if}\ \  t_1\not=(b,b)\\
(b-1)\frac{1}{1+m}\ \  &\text{if} \ \ t_1= (b,b).
\end{cases}
\end{align*}

To verify A2-A5 is the same as to verify the following inequalities, which are obviously true:
\begin{align*}(b-1)\frac{1}{1+m}&\geq 0\geq 0,\\
              b-1&\geq (b-1)\frac{1}{1+m}\geq(b-1)\frac{1}{1+m},\\
                          2(b-1)&\geq (b-1)\frac{1}{1+m}\geq 0,\\
b-1&\geq 0 \geq 0.
\end{align*}

\noindent{\textbf{Case C.}} $t_{-1}=( b^m 1^{n-m-1}, 1^{n-1})$, $m\geq 1$.

Symmetric to Case B, we have
\begin{align*} q_1(t_1, t_{-1})=
\begin{cases}
(0,0) \ \ \ \ &\text{if} \ \ t_1=(1,1) \\
(0,1)\ \  &\text{if} \ \ t_1=(1,b) \\
(\frac{1}{1+m}, \frac{1}{1+m})\ \  &\text{if} \ \ t_1=  (b,1)\\
(1,1)\ \  &\text{if} \ \ t_1=  (b,b);
\end{cases}
\end{align*}
\begin{align*}
u_1(t_1, t_{-1})=
\begin{cases}
0 \ \ &\text{if}\ \  t_1\not=(b,b)\\
(b-1)\frac{1}{1+m}\ \  &\text{if} \ \ t_1= (b,b).
\end{cases}
\end{align*}

To verify A2-A5 is equivalent to verifying the following inequalities, which are obviously true:
\begin{align*}b-1&\geq 0\geq 0,\\
              b-1&\geq (b-1)\frac{1}{1+m}\geq 0,\\
              2(b-1)&\geq (b-1)\frac{1}{1+m}\geq 0,\\
0&\geq 0 \geq -(b-1).
\end{align*}

We have so far treated all the cases when $t_1$ has at least one cheap item.  We divide the rest by whether $t_{-1}$ has an $i$ with $t_i=(b,b)$. Let $\ell(t_{-1})$ be the number of such $i$'s.

\noindent{\textbf{Case D.}}   $\ell(t_{-1})=0$;  $t^1_{-1}$ has $m$ $b$'s and  $t^2_{-1}$ has $m'$ $b$'s where $m, m'\geq 1$.

By the definition of $M_{D, \delta}$, it is easy to verify that
\begin{align*} q_1(t_1, t_{-1})=
\begin{cases}
(0,0) \ \ \ \ &\text{if} \ \ t_1=(1,1) \\
(0, \frac{1}{1+m'})\ \  &\text{if} \ \ t_1=(1,b) \\
(\frac{1}{1+m}, 0)\ \  &\text{if} \ \ t_1=  (b,1)\\
(1,1)\ \  &\text{if} \ \ t_1=  (b,b);
\end{cases}
\end{align*}
\begin{align*}
u_1(t_1, t_{-1})=0\ \ \  &\text{for all $t_1$}.
\end{align*}

To verify A2-A5 is equivalent to verifying the following inequalities, which are obviously true:
\begin{align*}(b-1)\frac{1}{1+m'}&\geq 0\geq 0,\\
              b-1&\geq 0 \geq 0,\\
              2(b-1)&\geq 0\geq 0,\\
(b-1)\frac{1}{1+m}&\geq 0 \geq -(b-1)\frac{1}{1+m'}.
\end{align*}

\noindent{\textbf{Case E.}}   $\ell(t_{-1})>0$.

By the definition of $M_{D, \delta}$, it is easy to verify that
\begin{align*} q_1(t_1, t_{-1})=
\begin{cases}
(0,0) \ \ \ \ &\text{if} \ \ t_1\not=(b,b) \\
(\frac{1}{1+\ell}, \frac{1}{1+\ell})\ \  &\text{if} \ \ t_1=  (b,b);
\end{cases}
\end{align*}
\begin{align*}
u_1(t_1, t_{-1})=0\ \ \  &\text{for all \  $t_1$}.
\end{align*}

To verify A2-A5 is equivalent to verifying the following inequalities, which are obviously true:
\begin{align*}0&\geq 0\geq 0,\\
              (b-1)\frac{1}{1+\ell}&\geq 0 \geq 0,\\
              2(b-1)\frac{1}{1+\ell}&\geq 0\geq 0,\\
0&\geq 0 \geq 0.
\end{align*}
We have shown that A2-A5 hold in all cases.  This completes the proof of Statement 1.  \qed

\vskip12pt

\noindent{\textbf{B. Proof of Statement 3}}
\vskip 8pt

We prove Statement 3 from Section 5.2.

\vskip 8pt
\noindent{\textbf{Statement 3.}}  $M_{B, \delta}$ is IR and BIC.

\vskip 8pt
\noindent{\emph{Proof}.}  Again, we assume $b>a=1$ without loss of generality. Mechanism $M_{B, \delta}$ is obviously IR, as the utility function for $M_{B, \delta}$ is defined to be always non-negative.  $M_{B, \delta}$ is also clearly BIC (and in fact DIC) if $b\in[v_3, \infty)$, as in this case $M_{B, \delta}=M_{D, \delta}$ which has been shown to be DIC in Appendix A.  Therefore, in our proof, we can assume that $b\in (1,v_3)$ and, in this situation, the parameters satisfy $\alpha\in \{0,1\}$ and $\beta=1$.

The BIC Conditions can be written as: for all $i, t_i, t'_i$,
\begin{align*} (t_i - t'_i)\cdot \bar{q}_i(t_i)\geq \bar{u}_i(t_i)-\bar{u}_i(t'_i) \geq (t_i - t'_i)\cdot \bar{q}_i(t'_i). \tag{A6}
\end{align*}

As $M_{B, \delta}$ is symmetric among items and among buyers, to show that $M_{B, \delta}$ is BIC, it is sufficient to prove the following subset of inequalities:
\begin{align*} (0, b-1)\cdot \bar{q}_1(1,b)&\geq \bar{u}_1(1,b)-\bar{u}_1(1,1)\\
                                               &\geq (0, b-1)\cdot \bar{q}_1(1,1)  \tag{A7} \\
               (b-1, 0)\cdot \bar{q}_1(b,b)&\geq \bar{u}_1(b,b)- \bar{u}_1(1,b)\\
                                               &\geq (b-1, 0)\cdot \bar{q}_1(1,b) \tag{A8} \\
             (b-1, b-1)\cdot \bar{q}_1(b,b)&\geq \bar{u}_1(b,b)- \bar{u}_1(1,1)\\
                                                 &\geq (b-1, b-1)\cdot \bar{q}_1(1,1)\tag{A9} \\
               (b-1, -(b-1))\cdot \bar{q}_1(b,1)&\geq \bar{u}_1(b,1)- \bar{u}_1(1,b)\\
                                                 &\geq (b-1, -(b-1))\cdot \bar{q}_1(1,b). \tag{A10}
\end{align*}

\noindent{\textbf{Fact A1.}} $\bar{q}_1(t_1)\geq \bar{q}_1(t'_1)$ if $t_1\geq t'_1$.

\vskip 8pt
\noindent{\emph{Proof}. The allocation function used by $M_{B, \delta}$ is consistent with the partial order on the types.  Thus, if $t_1\geq t'_1$,  we have $q_1(t_1, t_{-1})\geq q_1(t'_1, t_{-1})$ for any $t_{-1}$. This implies $\bar{q}_1(t_1)\geq \bar{q}_1(t'_1)$. \qed

\vskip 8pt
\noindent{\textbf{Fact A2.}}
$\bar{u}_1(1,b)-\bar{u}_1(1,1)=(0, b-1)\cdot \bar{q}_1(1,1)$, \ and\\
\hbox{}\ \ \ \ \ \ \ \ \ \ \ \ \ \  $\bar{u}_1(b,b)-\bar{u}_1(1,b)=(b-1, 0)\cdot \bar{q}_1(1,b)$.

\vskip 8pt
\noindent{\emph{Proof}. As by definition $u_1((1,1), t_{-1})=0$ for all $t_{-1}$, we have $\bar{u}_1(1,1)=0$.  We also have, by definition of $u_1$ for $M_{B,\delta}$,
\begin{align*}
\bar{u}_1(1,b)&=\sum _{t_{-1}}Pr\{t_{-1}\}u_1((1,b), t_{-1})\\
              &=p^{2n-2}\cdot u_1((1,b), (1,1)^{n-1})\\
              &=p^{2n-2}\cdot \frac{\alpha}{n}(b-1). \tag{A11}
\end{align*}

As $\bar{q}_1(1,1)=p^{2n-2}\cdot (\frac{\alpha}{n}, \frac{\alpha}{n})$, we have proved
\begin{align*}\bar{u}_1(1,b)-\bar{u}_1(1,1)=p^{2n-2}\cdot \frac{\alpha}{n}(b-1)=(0, b-1)\cdot \bar{q}_1(1,1).
\end{align*}
This proves the first equation in Fact A2. We now prove the other equation in Fact A2.  Let $T_1^j$ be the set of profiles $t_{-1}$ such that $t_{-1}^j$ is cheap (all 1's) while the other column has at least one $b$ appearing in it.  For each $t_{-1}$, let $\tilde{t}_{-1}$ be the type obtained by switching the two columns (i.e. interchanging items 1 and 2).  If $t_{-1}\in T_1^1$, clearly $\tilde{t}_{-1}\in T_1^2$, and vice versa.  We claim the following is true:

\noindent{\textbf{Claim A1.}} For any $t_{-1}\in T_1^1$,
\begin{align*}u_1((b,b), t_{-1})+u_1((b,b), \tilde{ t}_{-1})= (b-1)\cdot q^1_1((1,b), t_{-1}). \tag{A12}
\end{align*}

From the definition of $M_{B, \delta}$, we have for any $t_{-1}\in T_1^1$,
\begin{align*}u_1((b,b), t_{-1})+u_1((b,b), \tilde{ t}_{-1})= 2\cdot\frac{1}{2}(b-1)\frac{1}{1+|I(t_{-1})|}.
\end{align*}
Note also that $q^1_1((1,v), t_{-1})=\frac{1}{1+|I(t_{-1})|}$ because of $M_{B, \delta}$'s allocation function. Eq. A12 follows, and the Claim A1 is proved.

Now observe that, from the definition of $M_{B, \delta}$,
\begin{align*} u_1((b,b), (1,1)^{n-1})&=(b-1)(\frac{\alpha}{n}+1), \tag{A13}\\
u_1((b,b), t_{-1})&=0  \ \ \text{for all}\ \ t_{-1}\not\in\{(1,1)^{n-1}\}\cup T^1_1\cup T^2_1, \tag{A14}\\
\text{and}\ \ \ \ \ \ \ \ \ \ \ \ q^1_1((1,b), t_{-1})&=
\begin{cases} 1 \ \ \  \text{if}\ \   t_{-1}=(1,1)^{n-1}\\
              0  \ \ \ \text{if}\ \   t_{-1}\not\in\{(1,1)^{n-1}\}\cup T^1_1. \tag{A15}
\end{cases}
\end{align*}

It follows from Eqs A11, A13, and A14 that
\begin{align*}\bar{u}_1(b,b)-\bar{u}_1(1,b)
&=\sum _{t_{-1}}Pr\{t_{-1}\}u_1((b,b), t_{-1})-p^{2n-2}\frac{\alpha}{n}(b-1)\\
&=Pr\{(1,1)^{n-1}\}(b-1)(\frac{\alpha}{n}+1)-p^{2n-2}\frac{\alpha}{n}(b-1)\\
&\hbox{}    \  \ \ \ \ \ \ \ \ \ \                          +\sum _{t_{-1}\in T^1_1\cup T^2_1}Pr\{t_{-1}\}u_1((b,b), t_{-1})\\
&= Pr\{(1,1)^{n-1}\}(b-1)+\sum _{t_{-1}\in  T^1_1}Pr\{t_{-1}\}(u_1((b,b), t_{-1})+u_1((b,b), \tilde{t}_{-1})).                                       \end{align*}
Using Claim A1 and Eq. A15, we obtain
\begin{align*} \bar{u}_1(b,b)-\bar{u}_1(1,b)
&=(b-1)\big[Pr\{(1,1)^{n-1}\}+ \sum _{t_{-1}\in T^1_1}Pr\{t_{-1}\}q^1_1((1,b), t_{-1})\big]\\
&=(b-1)\bar{q}^1_1(1,b).
\end{align*}
This proves Fact A2.  \qed

From Fact A1, we have $\bar{q}_1(1,b) \geq \bar{q}_1(1,1)$ and $\bar{q}_1(b,b) \geq \bar{q}_1(1,b)$.  These inequalities and Fact A2 imply Eqs. A7 and A8.  To prove A9, we observe that Eqs. A7 and A8 together with  Fact A1 imply that
\begin{align*} (0, b-1)\cdot \bar{q}_1(b,b)&\geq \bar{u}_1(1,b)-\bar{u}_1(1,1)\\
                                               &\geq (0, b-1)\cdot \bar{q}_1(1,1)  \\
               (b-1, 0)\cdot \bar{q}_1(b,b)&\geq \bar{u}_1(b,b)- \bar{u}_1(1,b)\\
                                              &\geq (b-1, 0)\cdot \bar{q}_1(1,1).
\end{align*}
Adding up the above inequalities, we obtain Eq. A9.

It remains to prove Eq. A10.  We first establish the following Fact:

\noindent{\textbf{Fact A3.}} $\bar{q}^1_1(1,b)\leq  \bar{q}^2_1(1,b)$, \  $\bar{q}^1_1(b,1)\geq  \bar{q}^2_1(b,1)$.

Observe that
\begin{align*} q^1_1((1,b), t_{-1})=
\begin{cases} 1= q^2_1((1,b), t_{-1})  \ \ &\text{if}\ \   t_{-1}=(1,1)^{n-1} \\
\frac{1}{1+|I(t_{-1})|}=q^2_1((1,b), t_{-1})    \ \  &\text{if}\ \   t_{-1}\in T^1_1                  \\
              0\leq q^2_1((1,b), t_{-1})   \ \        &\text{otherwise}.
\end{cases}
\end{align*}
This implies $\bar{q}^1_1(1,b)\leq  \bar{q}^2_1(1,b)$.  The other inequality in Fact A3 follows by symmetry.  We have proved Fact A3. \qed

By symmetry $\bar{u}_1(b,1)-  \bar{u}_1(1,b)=0$.  Thus, to prove Eq. A10, it is equivalent to proving
\begin{align*}(b-1)(\bar{q}^1_1(b,1)- \bar{q}^2_1(b,1))\geq 0 \geq (b-1)(\bar{q}^1_1(1,b)- \bar{q}^2_1(1,b)).
\end{align*}
But this is true by Fact A3.  This proves  Eq. A10, and completes the proof of Statement 3.
\end{document}